\documentclass[authoryear,preprint,review,12pt]{elsarticle}

\usepackage{amssymb}
\usepackage{epsfig}
\usepackage{subfigure}
\usepackage{graphicx}
\usepackage{rotating}
\def\astrobj#1{#1}

\journal{New Astronomy}

\begin{document}

\begin{frontmatter}





\title{Detailed photospheric abundances of \astrobj{28 Peg} and \\ \astrobj{HD 202240}\tnoteref{t1}}
\tnotetext[t1]{Based on observations made at the T\"{U}B{\.I}TAK National Observatory, Turkey (Program ID 09BRTT150-477-0).}


\author{Asl{\i} Elmasl{\i}}

\author{\c{S}eyma \c{C}al{\i}\c{s}kan}

\author{Tolgahan K{\i}l{\i}\c{c}o\u{g}lu\corref{cor}}
\ead{tkilicoglu@ankara.edu.tr}

\author{K\"{u}bra\"{o}zge \"{U}nal}

\author{Yahya Nasolo}

\author{and Berahitdin Albayrak}

\cortext[cor]{Corresponding author. Tel.: +90 312 212 67 20; fax +90
312 223 23 95}

\address{Department of Astronomy and Space Sciences, Ankara University, 06100, Tando\u{g}an, Ankara, Turkey}

\begin{abstract}
The atmospheric parameters and chemical abundances of two neglected A-type stars, \astrobj{28 Peg} and \astrobj{HD 202240}, were derived using high resolution spectra obtained at the T\"{U}B{\.I}TAK National Observatory. We determined the photospheric abundances of eleven elements for \astrobj{28 Peg} and twenty for \astrobj{HD 202240}, using equivalent-width measurement and spectral synthesis methods. Their abundance patterns are in good agreement with those of chemically normal A-type stars having similar atmospheric parameters. We pinpoint the position of these stars on the H-R diagram and estimate their masses and ages as; $2.60\pm0.10$~M$_\odot$ and $650\pm50$~Myr for \astrobj{28 Peg} and $4.50\pm0.09$~M$_\odot$ and $150\pm10$~Myr for \astrobj{HD 202240}. To compare our abundance determinations with those of stars having similar ages and atmospheric parameters, we select members of open clusters. We notice that our target stars exhibit similar abundance patterns with these members.

\end{abstract}

\begin{keyword}
stars:individual:\astrobj{28 Peg} -- stars:individual:\astrobj{HD 202240}-- stars:abundances -- technique:spectroscopic
\end{keyword}
\end{frontmatter}


\section{Introduction}
\label{intro}
The precise elemental abundances of normal A-type stars provide information about chemical structure of their photospheres and give an idea of their evolutionary status. 

The spectral type of \astrobj{28 Peg} (\astrobj{HD 210516}, \astrobj{HR 8459}, \astrobj{BD+20 5093}) was classified as A3III by \citet{cowleyetal69}. Even though the star has different radial velocity values ($7.8~\mathrm{km~s^{-1}}$ \citep{shajnandalbitzky32}, and $10.46, 11.38$, $12.16~\mathrm{km~s^{-1}}$ \citep{kunzli98}) throughout the literature, \citet{kunzli98} noted that the star is non-variable. \citet{abtmorrell95} derived the $\mathrm{v\sin i}$ of \astrobj{28 Peg}, using Mg~II $4481$\,{\AA} line, as $40~\mathrm{km~s^{-1}}$. \citet{royeretal02} also reported its  $\mathrm{v\sin i}$ to be $49~\mathrm{km~s^{-1}}$.

\astrobj{HD 202240} (\astrobj{HR 8120}, \astrobj{BD+36 4470}) was classified as F0III \citep{cowleyetal69}. Its radial velocity values are given as $12.8~\mathrm{km~s^{-1}}$ \citep{harper37} and $13.8~\mathrm{km~s^{-1}}$ \citep{wilson53}. \citet{abtmorrell95} calculated the $\mathrm{v\sin i}$ of \astrobj{HD 202240}, using Mg~II $4481$\,{\AA} line, as $18~\mathrm{km~s^{-1}}$. The rotational velocity of \astrobj{HD 202240} is also given by \citet{royeretal02} as $26~\mathrm{km~s^{-1}}$. The first chemical abundance analysis of \astrobj{HD 202240} was carried out by \citet{kurtz76}, who used spectra having a dispersion of $8-10~\mathrm{{\AA}~mm^{-1}}$.

The aim of this paper is to perform abundance analysis of two normal A-type stars. The observations are described briefly in Section~\ref{obsdata}. The details of atmospheric parameters and abundance analysis are given in Section~\ref{atmpar} and Sections~\ref{abunana}. Finally, we present the results and conclusion in Section~\ref{redis}.

\section{Observation}
\label{obsdata}
The high resolution ($R\sim40,000$) spectra of \astrobj{28 Peg} and \astrobj{HD 202240} were obtained using the Coud\'{e} Echelle Spectrograph  
attached to the $1.5$\,m Russian-Turkish Telescope at T\"{U}B{\.I}TAK National Observatory. 
These spectra covering a wavelength range of 3900 to 7900\,{\AA}, were acquired on the $14^{th}$ of September and the $23^{rd}$ of December 2010.   
The observation run and data reduction procedures follow the same routine as described in \citet{caliskan05}. We co-added the two consecutive spectra of each star to achieve a higher signal-to-noise ratio ($S/N$). The properties and observation log of each target star are listed in Table~\ref{observation}.

\section{Atmospheric Parameters}
\label{atmpar}

We estimated the initial atmospheric parameters ($T_{\mathrm{eff}}$ and $\log~g$) of \astrobj{28 Peg} from the Str\"{o}mgren photometric data \citep{hauckmermilliod98} using the calibration of \citet{napiwotzkietal93}. As for the initial $T_{\mathrm{eff}}$ and $\log~g$ of \astrobj{HD 202240}, we used the Geneva colours of \citet{rufener76} with the calibration of \citet{kunzli97}. The adopted parameters are listed in Table~\ref{atmparam}. Using these parameters, we computed the initial model atmospheres for each star with ATLAS9 code \citep{kurucz93a,kurucz05,sbordoneetal04}.  

\begin{figure}
\centering
\begin{subfigure}
  \centering
  \includegraphics[trim=0mm 16mm 0mm 0mm, clip=true,width=10cm]{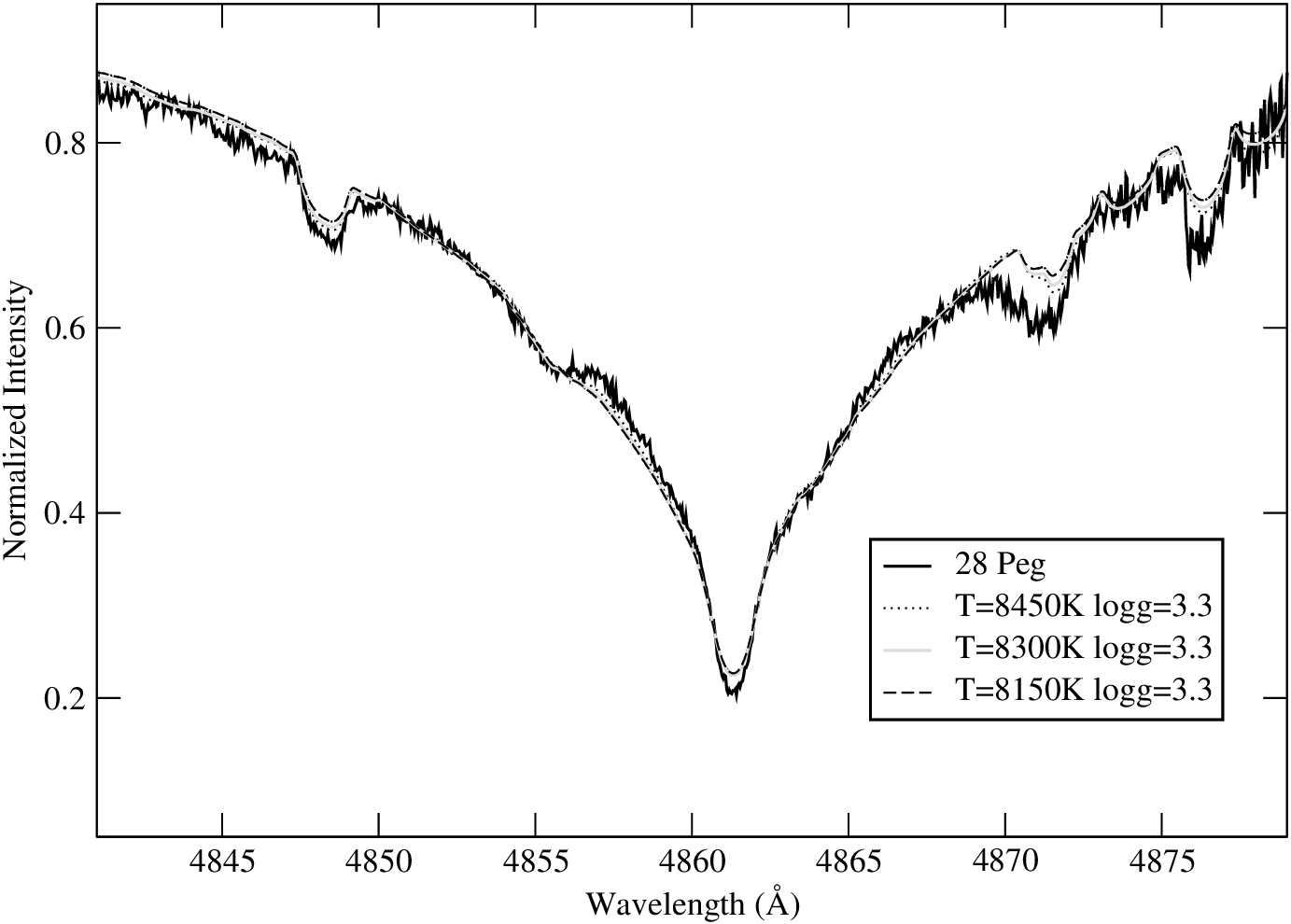}
  \label{cphbeta}
\end{subfigure}
\begin{subfigure}
  \centering
  \includegraphics[trim=0mm 0mm 0mm 0mm, clip=true,width=10cm]{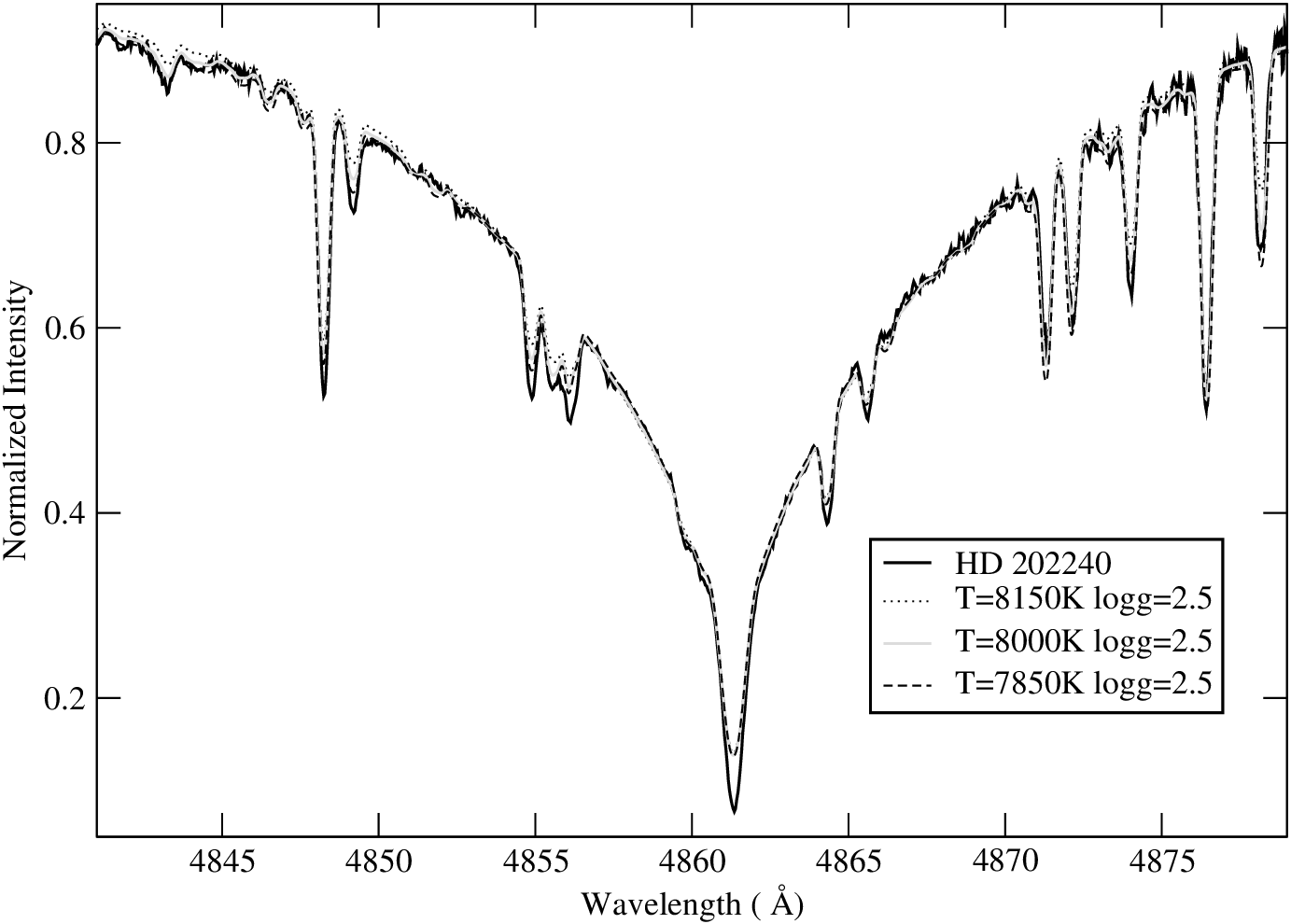}
  \label{hdhbeta}
\end{subfigure}
\caption{Comparison between the observed and synthetic H$_{\beta}$ line profiles of \astrobj{28 Peg} (top fig.) and \ \astrobj{HD 202240} (bottom fig.). 
The gray lines show the determined atmospheric parameters from this study, the dotted ($+150$\,K) and dashed ($-150$\,K) lines represent the uncertainties
 in $T_{\mathrm{eff}}$.}
\label{hbeta}
\end{figure}

These photometrically specified atmospheric parameters ($T_{\mathrm{eff}}$, $\log~g$) were then derived more precisely using traditional spectroscopic methods ($T_{\mathrm{eff}}$ from the excitation equilibrium of Fe~I lines and $\log~g$ from ionisation equilibrium of Fe~I/II). We also checked these parameters by generating synthetic H$_{\beta}$ profiles with SYNTHE code \citep{kurucz93b,kurucz05} and fitting these profiles to the observed ones, as presented in Figure~\ref{hbeta}. For the microturbulent velocities ($\xi$), we used the balance between equivalent-widths (hereafter EQWs) and abundances derived from individual Fe~I lines. The atmospheric parameters for each star are given in Table~\ref{atmparam}.

\section{Abundance Analysis}
\label{abunana}

In order to identify the absorption lines in the spectra, we used two atomic databases; Kurucz line database\footnote{http://kurucz.harvard.edu.tr} and Vienna Atomic Line Database \citep[VALD,][]{piskunovetal95,kupkaetal99,ryabchikovaetal99}. The EQWs were measured by fitting gaussian profiles to the observed lines. 
The WIDTH9 code \citep{kurucz05, sbordoneetal04}, based on ATLAS9 model atmospheres assuming line formation in LTE, was used to determine the abundances of each atomic species. EQWs greater than $190$\,m{\AA} were not used in any calculations.

The atomic data for lines affected by hyperfine splitting (HFS) were also taken from Kurucz line database. We then determined the abundances from these lines using synthetic spectra produced by SYNTHE code \citep{kurucz93b, kurucz05}. The synthetic spectra were convolved with the broadening effects due to the instrumental profile and the macroturbulent velocity. We justified the abundance value for each line until the observed and synthetic line profiles matched.
 
All derived elemental abundances within their uncertainties for each star are given in Table~\ref{abundance}. The total errors were calculated from the propogation of uncertainties in $T_{\mathrm{eff}}$, $\log~g$, and $\xi$ as given in \citet{caliskan05}.

\section{Results and Conclusion}
\label{redis}

\begin{figure}
\centering
 \includegraphics[trim=0mm 0mm 0mm 0mm, clip,width=15cm]{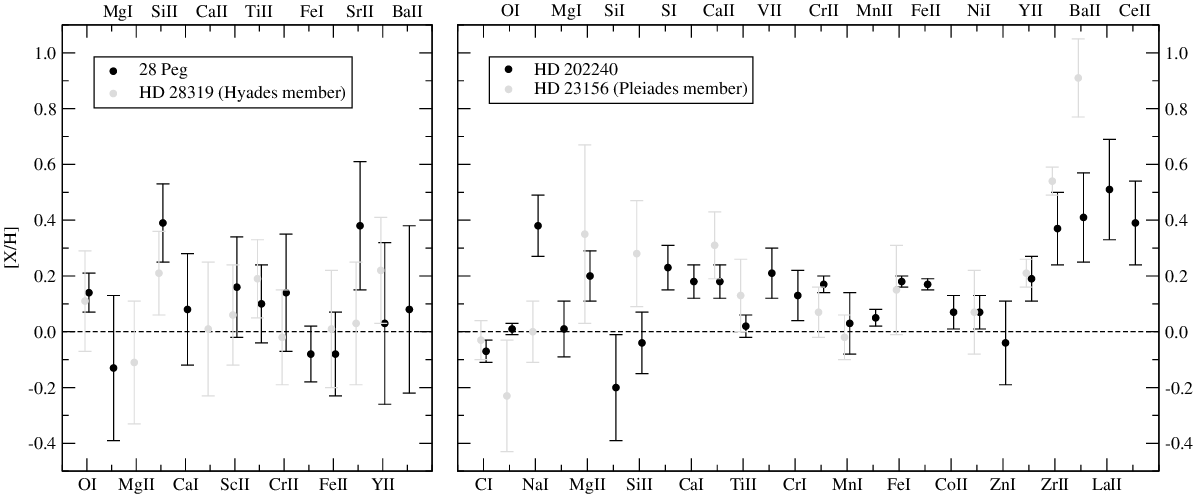}
  \caption{Comparison of chemical abundances between our target stars with their analogical cluster members. The solar abundances are from \citet{grevessesauval98} for all objects. The error bars are total uncertanties.}
 \label{abu}
\end{figure}

This is the first chemical abundance analysis of \astrobj{28 Peg} and \astrobj{HD 202240} based on high resolution spectra. The results indicate that the ions of both stars are slightly overabundant relative to the Sun, with a few exceptions that can be seen in Figure~\ref{abu}. These exceptions are; 
the ${\mathrm{[Si/H]}}$ and ${\mathrm{[Sr/H]}}$ abundances which are high as 0.38 in the atmosphere of \astrobj{28 Peg} and the abundances of the heavy elements ${\mathrm{[Ba/H]}}$, ${\mathrm{[La/H]}}$, ${\mathrm{[Zr/H]}}$, and ${\mathrm{[Ce/H]}}$ that are about 0.4 for \astrobj{HD 202240}. 
These are typical patterns for normal A-type stars as indicated in \citet{adelmanunsuree07}.

\begin{figure}
  \centering
  \includegraphics[trim=0mm 0mm 0mm 0mm, clip,width=10cm]{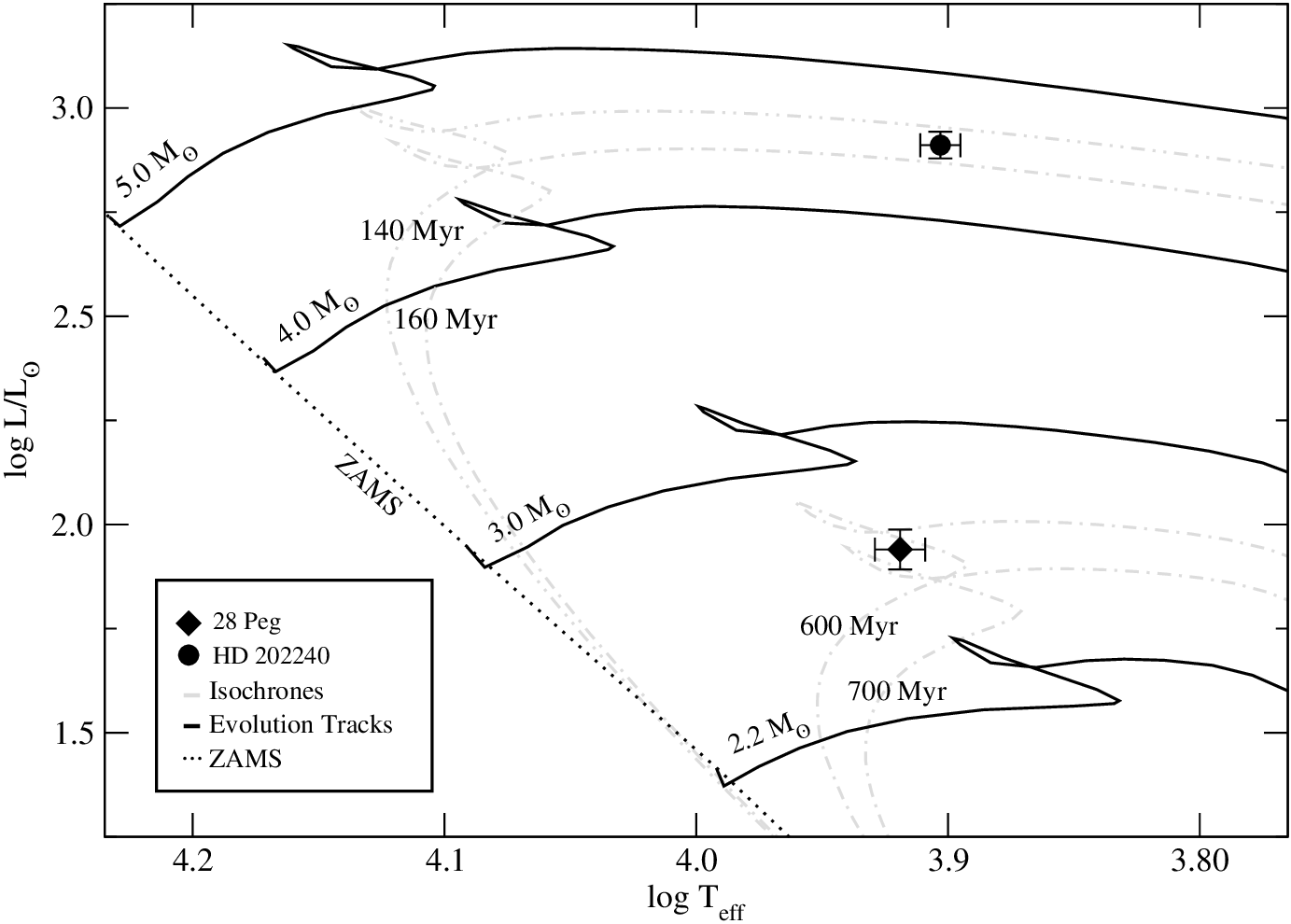}
  \caption{The position of \astrobj{28 Peg} and \astrobj{HD 202240} on the H-R diagram.  
Evolutionary tracks are from \citet{salasnichetal00} and isochrones from \citet{bressan12}.}
  \label{hr}
\end{figure}

The parameters in Table~\ref{hrt} were used to derive the luminosity of each star. We then plotted them on the H-R diagram as given in Figure~\ref{hr}, along with the evolutionary tracks with solar metallicity from \citet{salasnichetal00}, as black solid lines for masses of $2.2$~M$_\odot$, $3.0$~M$_\odot$, $4.0$~M$_\odot$, and $5.0$~M$_\odot$. Taking into account these evolutionary tracks, we note that \astrobj{28 Peg} and \astrobj{HD 202240} are giants with masses of $2.60\pm0.10$~M$_\odot$ and $4.50\pm0.09$~M$_\odot$, respectively. We also estimated the ages by using four isochrones ($140$, $160$, $600$, and $700$~Myr, shown as gray dash-dot lines in Figure~\ref{hr}) with solar metallicity taken from \citet{bressan12}. The age determined for \astrobj{28 Peg} is $650\pm50$~Myr and $150\pm10$~Myr for \astrobj{HD 202240}. 

These estimated ages of the stars allowed us to compare their abundance pattern with those of other stars of similar age and atmospheric parameters. For this comparison, we selected HD 28319 ($T_{\mathrm{eff}}$=7950 K, $\log~g$=3.70, age=625 Myr from \citep*{gebran10}) from the Hyades and HD 23156 ($T_{\mathrm{eff}}$=7940 K, $\log~g$=4.23, age=100 Myr \citep*{gebran08}) from the Pleiades open clusters. The comparison of chemical abundances between our target stars with corresponding cluster members showed that there is no significant difference, as shown in Figure~\ref{abu}. A few inconsistency, however, exist for only HD 202240. The Ba abundance of \astrobj{HD 202240} is lower than that of \astrobj{HD 23156}. This difference is not extraordinary since the abundance of Ba has a large star-to-star variation among A-type stars \citep{gebran08}. There is also a discrepancy for Na abundance. The overabundance of Na for HD 202240 most likely arises from non-LTE effects \citep{takeda08}. The difference in derived abundances between our and \citet{gebran08}'s study may also be due to the difference in selected Na I lines. Even though our target stars and comparison cluster members do not have any common origin, diffusion mechanisms seems to be responsible for these chemical similiarities. Additional spectral studies of normal A-type stars will help to understand the distrubition of the element in their atmospheres and evolutionary status. 

\bigskip
{\bf \noindent Acknowledgements}
\\ 
This work was supported by The Scientific and Technological Research Council of Turkey 
(T\"{U}B\.{I}TAK), the project number of 112T119. 









\begin{table}
  \caption[]{The properties and observation log of \astrobj{28 Peg} and \astrobj{HD 202240}.}
  \label{observation}
  \centering
  \tiny
  \begin{tabular}{r c c c c c c c c c}
    \hline
    \noalign{\smallskip}
    Star name & RA &  DEC & HJD & Exposure time& $\mathrm{v_{helio}}$ & $\mathrm{v\sin i}$&$S/N$ \\
&[$\mathrm{h}$~$\mathrm{m}$~$\mathrm{s}$]&[$^\circ$~$^\prime$~$^{\prime\prime}$]&[day]&$\mathrm{[s]}$&
$\mathrm{[km~s^{-1}]}$&$\mathrm{[km~s^{-1}]}$&[@$5000$\,{\AA}]\\
    \noalign{\smallskip}
    \hline
    \noalign{\smallskip}
\astrobj{28 Peg}  &$22~10~30.18$ &$+20~58~40.74$&$2455464.4633$&2$\times4500$ &$12.26\pm2.6$&$52$&$260$ \\    
\astrobj{HD 202240}&$21~13~26.42$ &$+36~37~59.75$&$2455555.1703$&2$\times4500$ &$-12.4\pm0.6$&$17$&$300$ \\            
    \noalign{\smallskip}
    \hline
  \end{tabular}
  \end{table}

\begin{table}
  \caption{The atmospheric parameters of the analyzed stars.}
  \label{atmparam}
  \centering
  \small
  \begin{tabular}{c c l r c c c} 
\hline
\noalign{\smallskip}
&\multicolumn{2}{c}{Photometric}&\multicolumn{4}{c}{Spectroscopic}\\ 
\noalign{\smallskip}
\hline
Star name &$T_{\mathrm{eff}}$ & $\log g$ &$T_{\mathrm{eff}}$& $\log g$ & $\xi$ &$\mathrm{[Fe/H]}$\\
&$\mathrm{[K]}$ &$\mathrm{[dex]}$ &$\mathrm{[K]}$&$\mathrm{[dex]}$&$\mathrm{[km~s^{-1}]}$&$\mathrm{[dex]}$\\
\noalign{\smallskip}
\hline
\noalign{\smallskip}
\astrobj{28 Peg}  & $8122$ &$3.38$& $ 8300$ & $3.30$&$3.00$ &$-0.12$\\
\astrobj{HD 202240}& $8092$ &$2.78$& $8000$ & $2.50$&$3.00$ &$ 0.18$\\
\noalign{\smallskip}
\hline
  \end{tabular}
  \end{table}

\begin{table}
   \caption[]{Derived elemental abundances with the standard random ($\sigma_{r}$) and total uncertainties ($\sigma_{tot}$) 
              of \astrobj{28 Peg} and \astrobj{HD 202240}. $\log\epsilon_{\odot}$ values are taken from \citet{grevessesauval98}.}
   \label{abundance}
   \centering
   \tiny
   \begin{tabular}{l c c r c c c c r c c c}
   \hline
   \noalign{\smallskip}
              &\multicolumn{5}{c}{\astrobj{28 Peg}}&\multicolumn{5}{c}{\astrobj{HD 202240}}\\
      \hline
      \noalign{\smallskip}      
      Species&$\log\epsilon_{\odot}$& $\log\epsilon$&$\mathrm{[X/H]}$&$\sigma_{r}$&$\sigma_{tot}$&N&$\log\epsilon$&$\mathrm{[X/H]}$&$\sigma_{r}$&$\sigma_{tot}$&N\\
\noalign{\smallskip}   
   \hline
   \noalign{\smallskip} 
C~I                    &8.52   & $\cdots$ & $\cdots$  & $\cdots$ & $\cdots$ & $\cdots$ &   8.45   & $-$0.07   &   0.03   &  0.04    &     8    \\      
O~I$\mathrm{^{SYN}}$   &8.83   &   8.97   &    0.14   &   0.05   &  0.07    &     4    &   8.84   &  0.01     &   0.02   &  0.02    &     3    \\      
Na~I                   &6.33   & $\cdots$ & $\cdots$  & $\cdots$ & $\cdots$ & $\cdots$ &   6.71   &  0.38     &   0.04   &  0.11    &     2    \\      
Mg~I                   &7.58   &   7.45   & $-$0.13   &   0.10   &  0.23    &     1    &   7.59   &  0.01     &   0.07   &  0.10    &     4    \\      
Mg~II                  &7.58   & $\cdots$ & $\cdots$  & $\cdots$ & $\cdots$ & $\cdots$ &   7.78   &  0.20     &   0.09   &  0.09    &     2    \\      
Si~I                   &7.55   & $\cdots$ & $\cdots$  & $\cdots$ & $\cdots$ & $\cdots$ &   7.52   & $-$0.03   &   0.17   &  0.19    &     3    \\      
Si~II                  &7.55   &   7.94   &    0.39   &   0.07   &  0.14    &     2    &   7.65   &  0.10     &   0.11   &  0.11    &     2    \\      
S~I                    &7.33   & $\cdots$ & $\cdots$  & $\cdots$ & $\cdots$ & $\cdots$ &   7.56   &  0.23     &   0.06   &  0.08    &     4    \\      
Ca~I                   &6.36   &   6.44   &    0.08   &   0.11   &  0.18    &     3    &   6.54   &  0.18     &   0.04   &  0.06    &     16   \\      
Ca~II                  &6.36   & $\cdots$ & $\cdots$  & $\cdots$ & $\cdots$ & $\cdots$ &   6.54   &  0.18     &   0.05   &  0.06    &     3    \\  
Sc~II$\mathrm{^{HFS}}$ &3.17   &   3.33   &    0.16   &   0.10   &  0.16    &     1    &  $\cdots$& $\cdots$  & $\cdots$ & $\cdots$ & $\cdots$ \\      
Ti~II                  &5.02   &   5.12   &    0.10   &   0.06   &  0.14    &     9    &   5.04   &  0.02     &   0.04   &  0.04    &     17   \\      
V~II$\mathrm{^{HFS}}$  &4.00   & $\cdots$ & $\cdots$  & $\cdots$ & $\cdots$ & $\cdots$ &   4.21   &  0.21     &   0.08   &  0.09    &     3    \\      
Cr~I                   &5.67   & $\cdots$ & $\cdots$  & $\cdots$ & $\cdots$ & $\cdots$ &   5.80   &  0.13     &   0.06   &  0.09    &     7    \\      
Cr~II                  &5.67   &   5.81   &    0.14   &   0.09   &  0.17    &     9    &   5.84   &  0.17     &   0.03   &  0.03    &     24   \\      
Mn~I$\mathrm{^{HFS}}$  &5.39   & $\cdots$ & $\cdots$  & $\cdots$ & $\cdots$ & $\cdots$ &   5.42   &  0.03     &   0.08   &  0.11    &     5    \\  
Mn~II$\mathrm{^{HFS}}$ &6.39   & $\cdots$ & $\cdots$  & $\cdots$ & $\cdots$ & $\cdots$ &   5.44   &  0.05     &   0.02   &  0.03    &     3    \\      
Fe~I                   &7.50   &   7.42   & $-$0.08   &   0.06   &  0.11    &     8    &   7.68   &  0.18     &   0.01   &  0.02    &     107  \\      
Fe~II                  &7.50   &   7.42   & $-$0.08   &   0.05   &  0.13    &    14    &   7.67   &  0.17     &   0.02   &  0.02    &     48   \\
Co~II                  &4.92   & $\cdots$ & $\cdots$  & $\cdots$ & $\cdots$ & $\cdots$ &   4.99   &  0.07     &   0.01   &  0.06    &     1    \\
Ni~I                   &6.25   & $\cdots$ & $\cdots$  & $\cdots$ & $\cdots$ & $\cdots$ &   6.32   &  0.07     &   0.03   &  0.06    &     9    \\      
Zn~I                   &4.60   & $\cdots$ & $\cdots$  & $\cdots$ & $\cdots$ & $\cdots$ &   4.56   & $-$0.04   &   0.01   &  0.15    &     1    \\      
Sr~II                  &2.97   &   3.35   &    0.38   &   0.10   &  0.23    &     1    &  $\cdots$& $\cdots$  & $\cdots$ & $\cdots$ & $\cdots$ \\      
Y~II$\mathrm{^{HFS}}$  &2.24   &   2.27   &    0.03   &   0.10   &  0.18    &     1    &   2.43   &  0.19     &   0.06   &  0.08    &     6    \\      
Zr~II                  &2.60   & $\cdots$ & $\cdots$  & $\cdots$ & $\cdots$ & $\cdots$ &   2.97   &  0.37     &   0.11   &  0.13    &     2    \\      
Ba~II$\mathrm{^{HFS}}$ &2.13   &   2.21   &    0.08   &   0.10   &  0.33    &     1    &   2.54   &  0.41     &   0.05   &  0.16    &     2    \\
La~II                  &1.17   & $\cdots$ & $\cdots$  & $\cdots$ & $\cdots$ & $\cdots$ &   1.68   &  0.51     &   0.01   &  0.18    &     1    \\      
Ce~II                  &1.58   & $\cdots$ & $\cdots$  & $\cdots$ & $\cdots$ & $\cdots$ &   1.97   &  0.39     &   0.01   &  0.15    &     1    \\ 
\noalign{\smallskip}
   \hline
   \end{tabular}
   \end{table}  
  
  \begin{table}
  \caption{The apparent magnitude in the V band, $m_{\mathrm{v}}$ (\citet{oja83} for \astrobj{28 Peg} and \citet{henden80} for \astrobj{HD 202240}), parallax, $\pi$ \citep{vanleeuwen07}, absolute magnitude, $M_{\mathrm{v}}$ (this study), bolometric correction, BC \citep{torres10}, luminosity, log$(L/L_{\odot})$ (this study), and logarithmic effective temperature, log $T_{\mathrm{eff}}$ (this study).}
  \label{hrt}
  \centering
  \tiny
  \begin{tabular}{c c c c c c c c} 
\hline
\noalign{\smallskip}
Star name &$m_{\mathrm{v}}$ & $\pi$ &$M_{\mathrm{v}}$ & BC & log$(L/L_{\odot})$ & log $T_{\mathrm{eff}}$ \\
          &$\mathrm{[mag]}$ &[$mas$]& $\mathrm{[mag]}$& $\mathrm{[mag]}$ & & $\mathrm{[K]}$\\
\noalign{\smallskip}
\hline
\noalign{\smallskip}
\astrobj{28 Peg}   & $6.46$ &$4.80\pm0.45$&$-0.136\pm0.011$&  $0.015$&$1.940\pm0.048$&$3.919\pm0.010$\\
\astrobj{HD 202240} & $6.08$ &$2.12\pm0.27$&$-2.576\pm0.005$&  $0.024$&$2.913\pm0.032$&$3.903\pm0.008$\\
\hline
\noalign{\smallskip}
  \end{tabular}
  \end{table}




\end{document}